**The Activity Reaction Core and Plasticity of Metabolic Networks**

E. Almaas[1,2], Z. N. Oltvai[3,*], and A.-L. Barabási[2,4]

[1]Microbial Systems Division, Biosciences Directorate, Lawrence Livermore National Laboratory, Livermore, California 94551, USA

[2]Center for Complex Networks and Department of Physics, University of Notre Dame, Notre Dame, Indiana 46556, USA

[3]Department of Pathology, University of Pittsburgh School of Medicine, Pittsburgh, Pennsylvania 15261, USA

[4]Center for Cancer Systems Biology, Dana Farber Cancer Institute, Harvard University, Boston, MA 02115, USA

[*] To whom correspondence should be addressed. Email: oltvai@pitt.edu

# ABSTRACT

**Background:**

Understanding the system level adaptive changes taking place in an organism in response to variations in the environment is a key issue of contemporary biology. Current modeling approaches such as the constraint-based flux balance analyses (FBA) have proved highly successful in analyzing the capabilities of cellular metabolism, including its capacity to predict deletion phenotypes, the ability to calculate the relative flux values of metabolic reactions and the properties of alternate optimal growth states.

**Methodology / Principal Findings:**

Here, we use FBA to thoroughly assess the activity of the *Escherichia coli*, *Helicobacter pylori*, and *Saccharomyces cerevisiae* metabolism in 30,000 diverse simulated environments. We identify a set of metabolic reactions forming a connected metabolic core that carry non-zero fluxes under all growth conditions, and whose flux variations are highly correlated. Furthermore, we find that the enzymes catalyzing the core reactions display a considerably higher fraction of phenotypic essentiality and evolutionary conservation than those catalyzing non-core reactions.

**Conclusions / Significance:**

Cellular metabolism is characterized by a large number of species-specific conditionally-active reactions organized around an evolutionary conserved always active metabolic core. Finally, we find that most current antibiotics interfering with the bacterial metabolism target the core enzymes, indicating that our findings may have important implications for antimicrobial drug target discovery.

# INTRODUCTION

Constraint-based modeling approaches, such as flux balance analysis (FBA) [1,2], have proved highly successful in analyzing the capabilities of cellular metabolism, including its capacity to predict deletion phenotypes, the ability to calculate the relative flux values of metabolic reactions and the properties of alternate optimal growth states in a wide range of simulated single-carbon source environmental conditions [3,4,5]. Recent analyses also indicate that the deletion phenotype of a substantial subset of metabolic enzymes is growth condition-dependent [6], arguing for a selective use of metabolic reactions in distinct environments.

# RESULTS

To thoroughly examine the utilization and relative flux rates of each metabolic reaction in a wide range of simulated environmental conditions, we sampled 30,000 randomly and uniformly chosen optimal growth conditions, as well as all single-carbon minimal medium conditions sufficient for growth, using FBA [1,2] on the reconstructed metabolic networks of *H. pylori*, *E. coli* and *S. cerevisiae* [7,8,9] (see Methods). We find that, when assuming optimal growth of the three microorganisms, their metabolism adopts to environmental changes through two distinct mechanisms. The more common is *flux-plasticity*, involving changes in the fluxes of already active reactions when the organism is shifted from one growth condition to another. For example, changing from glucose- to succinate-rich media alters the flux of 264 *E. coli* reactions by more than 20%. Less commonly, environmental changes can also induce *structural plasticity*, resulting in changes in the metabolism's active wiring diagram, activating previously zero-flux reactions and inhibiting previously active pathways. For example, when shifting *E. coli* cells from glucose- to succinate-rich media, 11 previously active reactions are turned off completely, while 9 previously inactive reactions are turned on.

The two types of response mechanisms described by flux and structural plasticity imply the possible existence of a group of reactions that are not subject to structural plasticity, being active under all environmental conditions. Indeed, some metabolic reactions were found to carry non-zero fluxes in *S. cerevisiae* under nine different growth conditions [6]. Yet, in itself this finding is compatible with a scenario in which the active reactions are randomly distributed: if typically a $q$ fraction of the metabolic reactions are active under a specific growth condition, for $n$ distinct conditions one can predict a nonzero overlap encompassing at least $q^n$ fraction of the reactions. Indeed, we find that as we increase the number of inspected conditions, $n$, the number of reactions carrying nonzero flux in each condition decreases rapidly. However, it eventually saturates at a constant value (Fig. 1a-c), identifying a group of reactions that have nonzero flux under *all* 30,000 simulated growth conditions. Specifically, we find that in *H. pylori* 138 of 381 (36.2%), in *E. coli* 90 of 758 (11.9%), and in *S. cerevisiae* 33 of 1172 (2.8%) metabolic reactions are always active (Fig. 1d,e). While these reactions respond to environmental changes only through flux-based plasticity, the rest of the reactions are only conditionally active, being turned on only in specific growth conditions and thus subject to both structural- and flux-based plasticity.

Figure 2 displays those metabolic reactions of *E. coli* that remain active in all 30,000 simulated growth conditions. The striking feature of this diagram is the fact that these reactions form a single connected cluster, encompassing each of the 90 reactions. This is not a unique feature of *E. coli*; in *H. pylori* all 138 reactions form a single cluster, and in *S. cerevisiae* all 33 reactions are connected. Given the relatively low number of always active reactions, the likelihood that they form a single large cluster by chance alone is vanishingly small, with P-values $< 1e^{-6}$ for *H. pylori* and *E. coli*, and $\sim 2e^{-6}$ for *S. cerevisiae*. Given the compact and clustered nature of the group of always-active reactions, we will refer to them collectively as the *metabolic core*.

The metabolic core contains two types of reactions: The first type consists of those that are essential for biomass formation under all environmental conditions (81 out of 90 reactions in *E. coli*), while the second type of reactions are required only to assure optimal metabolic performance (Table S1). In case of the inactivation of the second type, alternate sub-optimal pathways can be used to ensure cellular survival. Interestingly, in the compact core of *S. cerevisiae*, all 33 reactions predicted to be indispensable for biomass formation under all growth conditions. Moreover, when assuming a 10% reduction in the growth rate (compared to optimal growth) the size and identity of the *E. coli* metabolic core remains largely unchanged, retaining 83 of the original 90 reactions, of which two are nonessential. This indicates that the concept of the metabolic core is valid under both optimal- and suboptimal growth, although with some difference in the identity of individual reactions (a detailed study on this issue will be published elsewhere). Of note, the metabolic core represents a subset of the minimal reaction sets [10,11] and the overlap of alternative (degenerate) minimal reaction sets [4,5]: The minimal reaction set of Ref. 11 contains the metabolic core in addition, however, to reactions necessary for the sustained growth on any chosen substrate, whereas the minimal reaction set of Ref. 5 consists of the 201 reactions that are always active in *E. coli* for all 136 aerobic and anaerobic single-carbon source minimal environments capable of sustaining optimal growth. The latter finding is in excellent agreement with Fig 1b (upper curve), which demonstrates that a low number of growth conditions will significantly overestimate the set of always active reactions.

To identify some of the factors that determine the size of the metabolic core, we note that the number of core metabolic reactions systematically decrease as we move from *H. pylori* to *S. cerevisiae* (Fig. 1e), the relative size of the core decreasing from 36.2% of all reactions in *H. pylori* to 11.9% in *E. coli* and 2.8% in *S. cerevisiae*. This trend can be explained as a collective network effect: the relatively small size of the *H. pylori* metabolic network leaves little flexibility for biomass production, requiring the continuous activity of a high fraction of the available metabolic reactions. Indeed, we find that on average approximately 61% of the *H. pylori* reactions are active in a given environmental condition. The larger number of metabolic reactions present in *E. coli* offers a higher degree of metabolic flexibility, allowing for a significant fraction of the biomass to be produced by alternate pathways. Indeed, the average utilization of the *E. coli* metabolic network in a given growth condition is only 35.3%. For *S. cerevisiae*, whose metabolic network size significantly exceeds that of both *H. pylori* and *E. coli*, there is an even higher metabolic flexibility and the activity of only 19.7% of the reactions are required in a typical environment.

The fact that the core reactions are active under all investigated environmental conditions suggests that they must play a key role in maintaining the metabolism's overall functional integrity. Therefore, the absence of individual reactions that are part of the metabolic core may lead to significant metabolic disruptions. Indeed, using genome-scale deletion phenotype data obtained in rich growth media [12,13], we find that 74.7% of those *E. coli* enzymes that catalyze core metabolic reactions (i.e., *core enzymes*) are essential, compared with the 19.6% lethality fraction characterizing the non-core enzymes. The gap between the core and non-core is significant for *S. cerevisiae* as well, for which essential enzymes catalyze 84% of the core reactions, whereas the average essentiality of the conditionally active enzymes is only 15.6% (Fig. 3b). Note, that the likelihood of a random concentration of so many essential enzymes in the core is extremely small, with P-values of $3.3e^{-23}$ and $9.0e^{-13}$ for *E. coli* and yeast, respectively. Assuming the presence of errors in the deletion phenotype data does not significantly change our findings (see Supplementary Material). Taken together, these results indicate that an organism's ability to adapt to changing environmental conditions to a large extent rests on the continuous activity of the metabolic core, regardless of the environmental conditions.

Intuitively, one could assume that the core represents a subset of high flux reactions characterizing the activity of metabolic networks [3]. Yet, our measurements indicate that on average the fluxes of the core and non-core reactions are highly comparable (Fig. 3c). Alternatively, we could also assume that the main biological role of the metabolic core is to ensure the continuous production of biomass under all growth conditions. In contrast, we find that 20 of the 51 metabolites (39%) considered necessary for biomass production in *E. coli* [8] are not produced by any of the core reactions; instead - in a growth-condition dependent fashion-, they are produced by various alternative metabolic pathways. The core, however, contains a large number of reactions for selected anabolic pathways, including those of membrane lipid-, cell envelope- and peptidoglycan biosynthesis pathways (see Supplementary Material). These core reactions represent network bottlenecks, being the only paths for the synthesis of certain biomass components. Therefore, it appears that the composition of the metabolic core is determined by two factors: those metabolic reactions that directly contribute to biomass production tend to be part of the metabolic core of the organism. However, this tendency is offset by network-induced redundancy: reactions or pathways, whose end products can be synthesized by at least two alternate pathways show environmental redundancy and structural plasticity, thus are eliminated from the core. Therefore, the more reactions a metabolic network possesses (Fig. 1d), the stronger is the network-induced redundancy, and the smaller is the core (Fig. 1e).

Given the important functional role played by the metabolic core in a given organism, one expects significant parts of the core to be conserved in different organisms. Indeed, the *E. coli* and *H. pylori* cores have 63 reactions in common, and 18 of the 33 core reactions in *S. cerevisiae* are present in both the *E. coli* and the *H. pylori* metabolic cores (Fig. 3a). Also, when considering enzyme orthologs among 32 divergent bacteria [13], we find that the metabolic core enzymes of *E. coli* display a high degree of evolutionary conservation, the average core enzyme having orthologues in 71.7% of the reference bacteria (P-value < 1e-6). In contrast, the conditionally active non-core enzymes have an evolutionary retention of only 47.7% [13]. Given the observed correlation between evolutionary retention and the essentiality of gene products [13], this difference may be a simple consequence of the high lethality fraction of the core enzymes. Yet, further analysis indicates that this is not the case: random selection of 90 enzymes with a 74.7% lethality ratio has an average evolutionary retention of only 63.4%. Taken together, these results indicate that major portions of the metabolic core have been conserved, displaying a higher evolutionary retention than the individual essentiality of its participating enzymes would indicate, suggesting that maintaining the core's integrity is a collective need of the organism.

The requirement for the continuous activity of the core reactions might have impacted the regulation of its catalytic enzymes as well. We would expect that the activity of core reactions, -for which only the flux magnitude needs to be modulated-, should display a higher degree of stability and a different regulatory control than the reactions outside of it. Evidence for such regulatory effects is provided by two measurements. First, we inspected the experimentally determined mRNA half-lives of the *E. coli* metabolic enzymes when cells were grown in Luria-Bertani (LB) medium [14]. We find that the distribution of mRNA half-life times for the core enzymes displays a shift to higher values, with the average half-life time of their mRNAs being 14.0 minutes compared with 10.5 minutes for the non-core mRNAs (P-value = 0.016). Furthermore, only 2% of the core enzymes have corresponding mRNA half-life times shorter than 5 minutes, compared to 23% for the ones catalyzing conditionally active non-core reactions (see Methods). Therefore, the continuous metabolic need for these reactions has apparently affected the mechanisms responsible for mRNA decay of their catalytic enzymes, providing a higher dynamic inertia under environmental changes. Additionally, for each enzyme-encoding operon we counted the number of activating and repressive regulatory links in the *E. coli* transcriptional-regulatory network [15,16]. As regulatory interactions are currently known for only 13 of the core enzyme-encoding operons, we considered the extended core, representing the set of 234 reactions that are active in more than 90% of the 30,000 simulated growth conditions. The results indicate that the fraction of repressive regulatory links in the extended core is 52.3%, while the fraction of activating interactions is only 35.7%, the remaining 12% representing regulatory links that can either activate or suppress the enzyme's mRNA synthesis rate. In contrast, for non-core enzyme-encoding operons there is no difference between the fraction of activating and repressing links, both representing 45% of the regulatory interactions. These results offer evidence for the co-evolution of the core metabolic- and its corresponding regulatory network: given the requirement for the continuously active non-redundant core, the regulatory mechanisms have shifted both the mRNA decay rate and the nature of the regulatory interactions, offering a higher regulatory stability for the core enzymes.

The finding that the core reactions form a single cluster would suggest that the activity of the participating reactions is highly synchronized, changes in the flux of one reaction affecting the flux of other reactions as well. Thus, we should be able to discover the core by inspecting the correlations between the activities of all metabolic reactions. To test this hypothesis, we calculated the flux correlation coefficient, $C_{ij}$, for each reaction pair in the *E. coli* metabolism [17] by inspecting the flux values of all reactions under each of the 30,000 simulated growth conditions. Using the $C_{ij}$ matrix as the metric of a hierarchical clustering algorithm [18], we observe the emergence of a group of reactions whose fluxes change simultaneously under environmental shifts. Interestingly, these highly correlated reactions significantly overlap with the metabolic core (Fig. 4a, and Supplementary Information), indicating that when the environmental conditions are altered, the fluxes of the core reactions are increased or decreased in a synchronized fashion. The same synchrony is observed when we inspect experimental mRNA data among the core reactions during changes in environmental conditions. Calculating the correlations between core and non-core enzymes using their corresponding mRNA copy number data (see Methods) for 41 experiments [19] we find that the correlations among the core enzymes' mRNA copy numbers are systematically higher than the observed correlations among their non-core counterparts (Fig. 4b), with an average correlation of <C>=0.23 for core enzymes and <C>=0.07 for non-core enzymes (P-value < 1e-4). Notably, this finding is different from the average correlation of mRNA expression found in 66 'correlated reactions sets' (groups from two to nine reactions turned on and off together), where the frequency with which a correlated reaction set is used does not affect its mRNA correlation [5].

**DISCUSSION**

Previous studies have firmly established that environmental changes induce flux plasticity, altering the flux levels of individual reactions (reviewed in Ref. 20). Similarly, the fact that metabolism displays structural plasticity, turning on and off some reactions as the growth conditions are altered, has been observed before [3,4,6]. Yet, our demonstration of a group of reactions predicted to be active in all environmental conditions and forming a connected metabolic core could substantially improve our understanding of the organization and utilization of metabolic networks. The emergence of the core represents a collective network effect, channeling the production of some indispensable biomass components to a few key reactions that cannot be replaced by alternative pathways under any environmental conditions. The collective origin of the core is supported by the observed changes in the core size. While the number of biomass components that the metabolism needs to produce is comparable in the three organisms (49 for *H. pylori*, 51 for *E. coli* and 44 for *S. cerevisiae*), the number of metabolic reactions contributing to them differs. The larger size of a metabolic network significantly increases its capacity and redundancy, decreasing the metabolic core's size. Therefore, the metabolic core contains reactions that are necessary for optimal cellular performance regardless of the environmental conditions, while the conditionally active metabolic reactions represent the different ways the cell is capable of optimally utilizing substrates from its environment.

The identification of the metabolic core has important practical implications, too: given the continuous activity and high degree of essentiality of the core enzymes, they represent potential targets for antimicrobial intervention. Specifically, while many bacterial and yeast gene products are essential, a high fraction of them are essential only in specific environments. For example, recent measurements indicate that 76% of the *S. cerevisiae* genes that are inactive in nutrient rich conditions are in fact not only active, but also essential in some other growth conditions [6]. Yet, an effective antimicrobial drug needs to be able to kill its target organism under all physiological conditions that it can exist. Thus pharmacological interventions targeting a specific pathway will not be effective in environments where the pathway is not needed and its corresponding enzyme is turned off. Instead, the most effective antimicrobials must target the activity of the core reactions, as their disruption will impact the microorganism's ability to function under all environmental conditions. Indeed, among the currently used antibiotics, fosfomycin and cycloserine act by inhibiting cell wall peptidoglycan-, while sulfonamides and trimethoprim inhibit tetrahydrofolate biosynthesis, both pathways being present both in the *H. pylori* and the *E. coli* core (Fig. 2). In addition, core pathways involved in the synthesis of Flavin adenine dinucleotide (FAD), 2-Dehydro-3-deoxy-D-oxtonate (KDO) and Lipopolysaccharide (LPS) are among currently explored potential antibiotic targets [21]. This indicates that the core enzymes essential for biomass formation, both for optimal and suboptimal growth (Table S1), may prove effective antibiotic targets given the cell's need to maintain their activity in all different conditions. The fact that not all such core reactions are shared by all bacteria offers the possibility to identify bacterium-specific drug targets. Finally, our results pertaining to the existence of the core and its characteristics are by no means limited to the three studied organisms, but the analysis can be carried over to all organisms for which high quality metabolic reconstructions is available. Given the large number of sequenced bacterial genomes, such studies could open new avenues for rapid *in silico* antimicrobial drug target identification.

**MATERIALS AND METHODS**

**Flux balance analysis (FBA)** Starting from the published stoichiometric matrices of the reconstructed *E. coli* MG1655, *H. pylori*, and *S. cerevisiae* metabolic networks [7,8,9], the steady state concentrations of all the internal metabolites of an organism satisfy

$$\frac{d}{dt}[A_i] = \sum_j S_{ij} v_j = 0, \qquad (1)$$

where $S_{ij}$ is the stoichiometric coefficient of metabolite $A_i$ in reaction $j$ and $v_j$ is the flux of reaction $j$. We use the convention that if metabolite $A_i$ is a substrate (product) in reaction $j$, $S_{ij} < 0$ ($S_{ij} > 0$). Any vector of fluxes $\{v_j\}$ which satisfies Eqn. (1) corresponds to a state of the metabolic network, and hence, a potential state of operation of the cell. Using linear programming, we choose the solution that maximizes biomass production of the respective organisms. Reactions that are never active, likely reflecting annotation errors or data incompleteness, are ignored in our flux analysis.

**Metabolic core identification** We model all possible cellular environments in the investigated FBA models, as described before [3]. Briefly, for each metabolic uptake reaction we fix the uptake rate to a randomly chosen value between 0 and 20 mmol/g DW/h before calculating the optimal fluxes for this configuration using linear programming. Each reaction is subsequently tagged as either active (non-zero flux) or inactive (zero flux). Since there are a very large number of possible combinations of the selected uptake rates, we repeat this process 30,000 times. We additionally calculate the optimal fluxes for all single carbon-source configurations on a minimal uptake medium consisting of unlimited ammonia, sulfate, phosphate, carbon dioxide, potassium and restricted oxygen for *E. coli* and *H. pylori*, and unlimited ergosterol and zymosterol as well for *S. cerevisiae*. For each reaction we assign a value, $q_i$, between zero and unity, representing the relative number of conditions for which reaction $i$ is active. The preliminary metabolic core is defined as the set of reactions that are active in all sampled conditions ($q_i=1$). We determine the final metabolic core by removing those reactions from the preliminary core that, when their flux rate is constrained to zero over the total set of initial conditions, leave the growth rate unchanged.

**Correlations and clustering** We evaluate the FBA flux correlations by calculating the Pearson coefficient for all possible reaction pair combinations in the 30,000 different simulated growth conditions. We subsequently group the reactions by employing the correlation values as the metric in a standard hierarchical average-group linkage clustering algorithm [22].

**Data analysis** The annotated metabolic FBA models specify the enzymes and genes catalyzing the various metabolic reactions. We represent the mRNA activity level or half-life time of a metabolic enzyme by averaging over the available experimental values, $E_k$, for all the catalyzing gene products, as $r_i = \frac{1}{N_i} \sum_{k=1,N_i} E_k$, where $N_i$ is the number of genes. We assess the essentiality of a reaction by considering it essential if at least one of its catalyzing enzymes (or gene products) is essential for the survival of the organism. See Supplementary Material for error analysis.

**Core and non-core essentiality** The deletion phenotype of *E. coli* enzymes was taken from Gerdes *et al* [13]. For non-lethal core reactions we cross-checked the deletion phenotype with the PEC database (http://www.shigen.nig.ac.jp/ecoli/pec/) as given in Table S2 of Gerdes *et al*, resulting in a total of 62 essential enzymes and 7 enzymes for which we could not determine the deletion phenotype. The deletion phenotype of the yeast enzymes was taken from the CYGD database [12].

**Acknowledgement**

We thank S. Mobashery and S. Vakulenko for discussions. Work at the University of Notre Dame and at the University of Pittsburgh was supported by the US Department of Energy, NIH and NSF.

**Competing Interest**. The authors have declared that no competing interests exist.

**Author contributions.** EA, ZNO, and ALB conceived the study. EA and ALB designed the study. EA performed all the computational work. EA, ZNO and ALB wrote the paper.

**FIGURE CAPTIONS**

**Figure 1. The emergence of the metabolic core.** The average relative size of the number of always active reactions as a function of the number of sampled conditions (black line) for **(a)** *H. pylori*, **(b)** *E. coli* and **(c)** *S. cerevisiae*. As the number of conditions increases, the curve converges to a constant denoted by the dashed line, identifying the metabolic core of an organism. The red line denotes the number of always active reactions if activity is randomly distributed in the metabolic network. The fact that it converges to zero indicates that the real core represents a collective network effect, forcing a group of reactions to be active in all conditions. **(d)** The number of metabolic reactions, and **(e)** the number of metabolic core reactions in the three studied organisms.

**Figure 2. The metabolic core of *E. coli*.** The figure shows all reactions that are found to be active in each of the 30,000 investigated external conditions. Metabolites that contribute directly to biomass formation [8] are colored blue, while core reactions (links) catalyzed by essential (non-essential) enzymes [13] are colored red (green) (black colored links denotes enzymes with unknown deletion phenotype). The blue dashed lines indicate multiple apparances of a metabolite, while links with arrows denote unidirectional reactions. Note that 20 of the 51 metabolites necessary for biomass synthesis are not present in the core, indicating that they are produced (consumed) in a growth condition-specific manner. See the Supplementary Material for the abbreviations of metabolites and a list of core reactions for *E. coli*, *H. pylori* and *S. cerevisiae*. The folate and peptidoglycan biosynthesis pathways are indicated by blue and brown shading, respectively, and the white numbered arrows denote current antibiotic targets inhibited by (1) sulfonamides, (2) trimethoprim, (3) cycloserine, and (4) fosfomycin. Note that a few reactions appear disconnected since we have surpressed the drawing of cofactors.

**Figure 3. Characterizing the metabolic cores.** **(a)** The number of overlapping metabolic reactions in the metabolic core of *H. pylori*, *E. coli* and *S. cerevisiae*. **(b)** The fraction of metabolic reactions catalyzed by essential enzymes in the cores (black) and outside the core in *E. coli* and *S. cerevisiae*. **(c)** The distribution of average metabolic fluxes for the core and the non-core reactions in *E. coli*.

**Figure 4. Correlations among *E. coli* metabolic reactions.** **(a)** We calculated the Pearson correlation using flux values from 30,000 conditions for each reaction pair before grouping the reactions according to a hierarchical average-linkage clustering algorithm. The values of the flux-correlation matrix go from negative one (red) through zero (white) to unity (blue). The horizontal color bar denotes if a reaction is member of the *core* (green) and the vertical if the enzymes catalyzing the reaction are essential (red). **(b)** Distribution of Pearson correlations in mRNA copy numbers from 41 experiments [14]. The correlations of the core reactions are clearly shifted towards higher values, with an average correlation coefficient of <C>=0.23 compared with the average non-core coefficient of <C>=0.07.

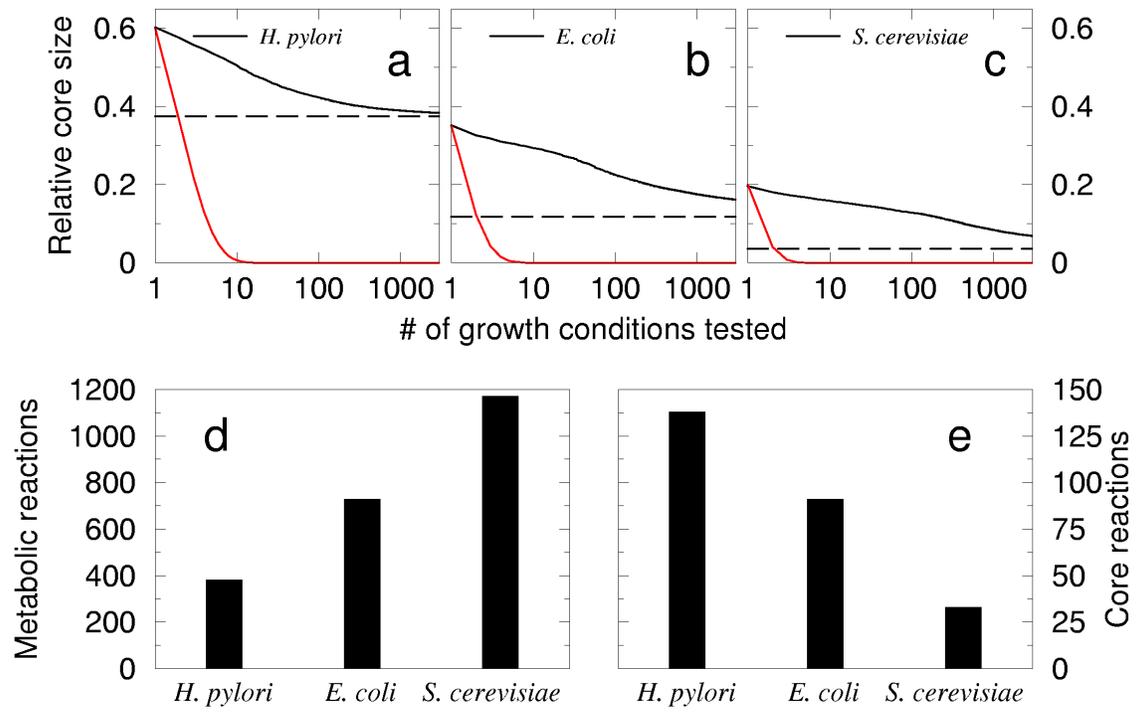

**Figure 1.**

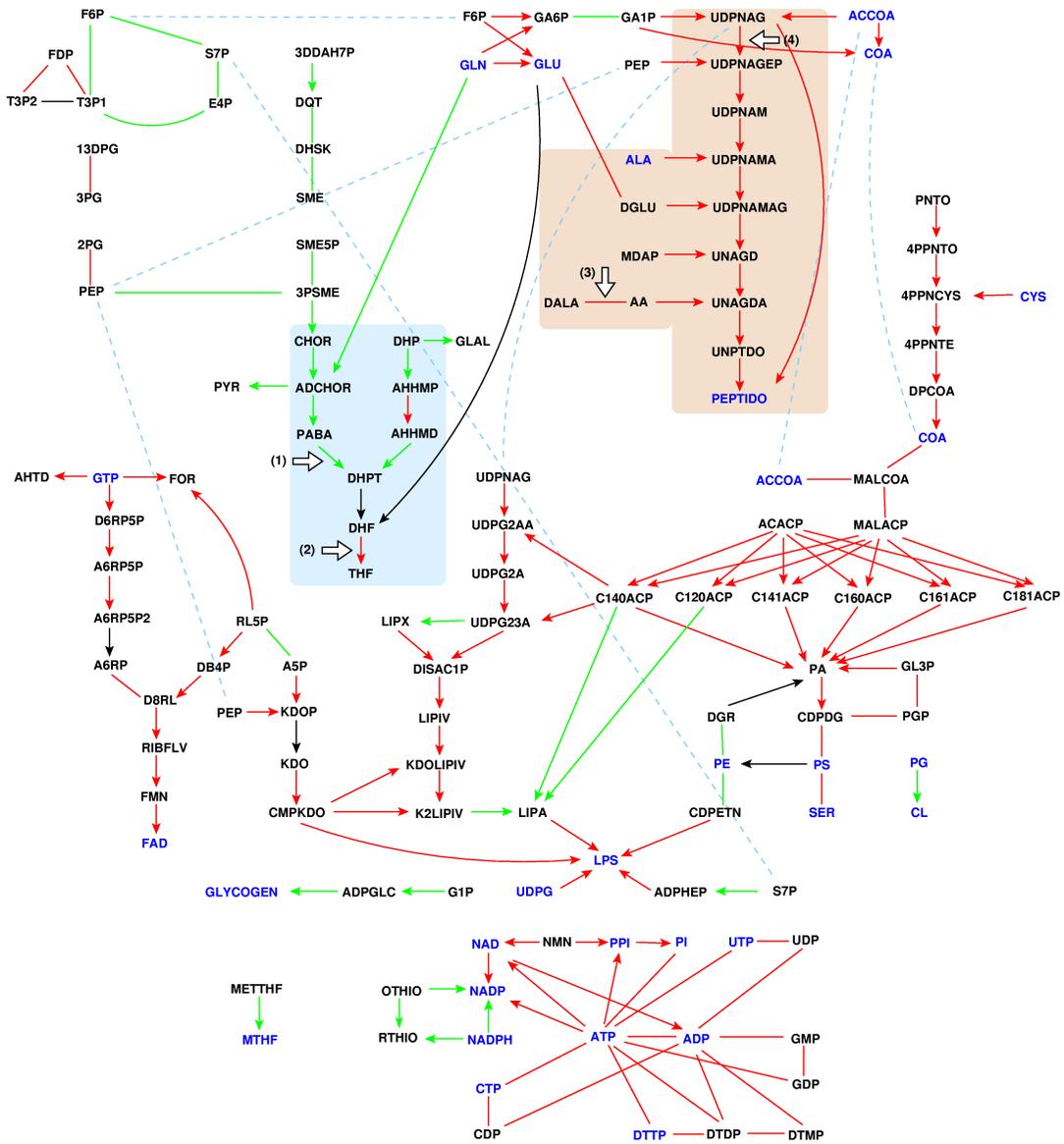

**Figure 2.**

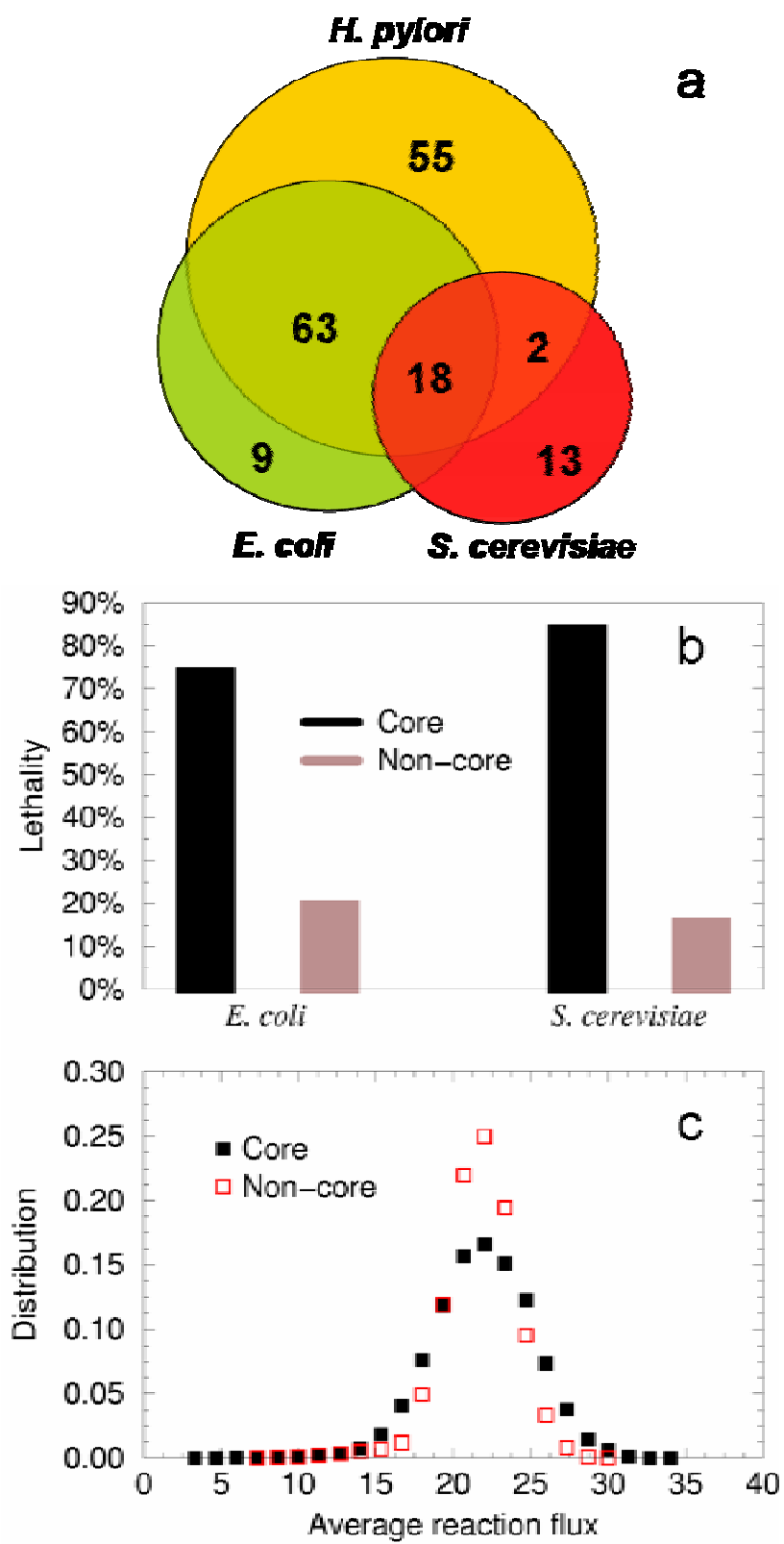

**Figure 3.**

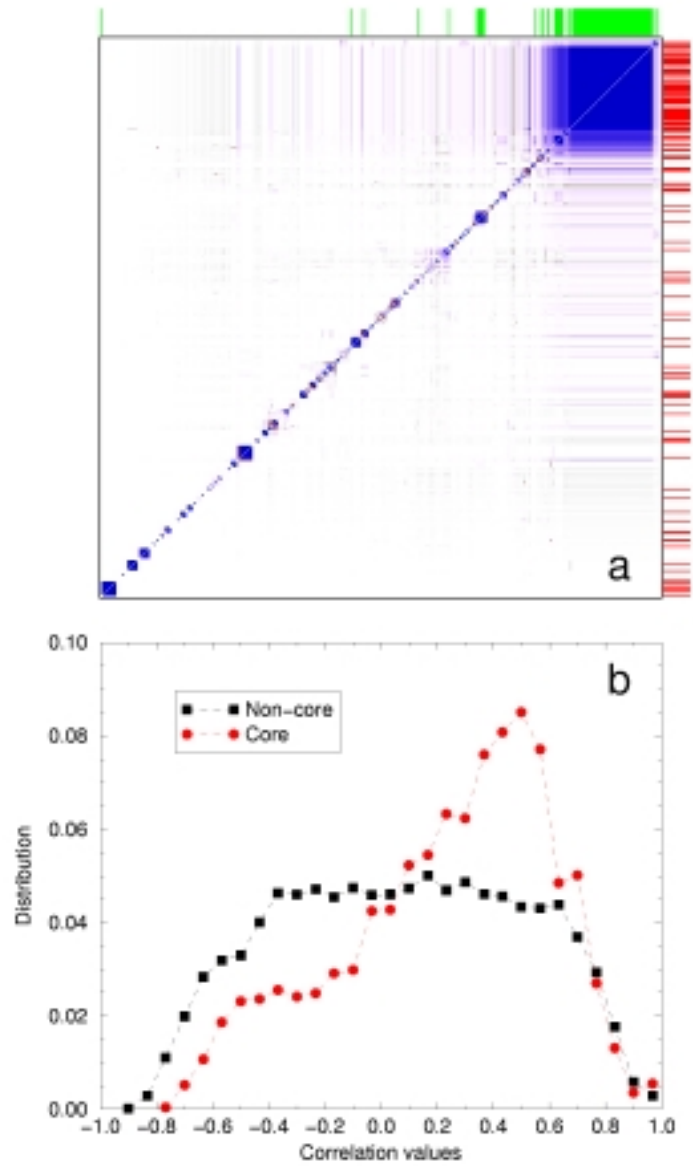

**Figure 4.**